\documentclass[aps,12pt]{revtex4}
\input{amssym}

\def\p {\partial}
\def\t {\tilde}
\def\be {\begin{equation}}
\def\ee  {\end{equation}}
\def\bea {\begin{eqnarray}}
\def\eea {\end{eqnarray}}

\begin{document}
\baselineskip=1.5em
%\draft

\title{Semiclassical states for quantum cosmology}

\author{Viqar Husain and Oliver Winkler}

\affiliation{\baselineskip=1.5em Department of Mathematics and Statistics,\\
University of New Brunswick, Fredericton, NB E3B 5A3, Canada.  \\
Perimeter Institute of Theoretical Physics\\
Waterloo, ON Canada.\\
EMail: husain@math.unb.ca, owinkler@perimeterinstitute.ca\\
\mbox{}}

\thispagestyle{empty}

%\date{July 22, 2006}

\begin{abstract}
\baselineskip=1.5em

In a metric variable based Hamiltonian quantization, we give a
prescription for constructing semiclassical matter-geometry states
for homogeneous and isotropic cosmological models. These
"collective" states arise as infinite linear combinations of
fundamental excitations in an unconventional "polymer" quantization.
They satisfy a number of properties characteristic of
semiclassicality, such as peaking on classical phase space
configurations. We describe how these states can be used to
determine quantum corrections to the classical evolution equations,
and to compute the initial state of the universe by a backward time
evolution.

\end{abstract}

%%%%%%%%%%%%%%%%%%%%%%%%%%%%%%%%%%%%%%%%%%%%%%%%%%%%%%%%%%%%

%\vs{1cm}

\maketitle

%\begin{multicols}{2}

\section{Introduction}

One of the foundational questions in cosmology is how a large
universe described effectively by classical physics emerges from a
small and highly quantum one. There are many facets to this
question, ranging from a "theory of initial conditions" for the
Wheeler-DeWitt equation to the origin of the vacuum state
responsible for the emergence of density fluctuations, to how such
fluctuations become classical. At present, there are no final
answers to these questions.

While the relation between a quantum system and its classical
counterpart has many facets, see \cite{landsman} for a nice overview
and \cite{kiefer} for a discussion in the context of quantum
gravity, the notion of semiclassical state plays an important role.
For a "standard" quantum system such as the harmonic oscillator
there is a well-developed notion of semiclassical state, namely the
"coherent state", characterized by properties such as minimum
uncertainty, peakedness on a classical configuration, and the
relationship to classical physics that arises via Ehrenfest
theorems. For quantum gravity the WKB approximation has been the
more common approach for exploring semiclassical physics, although
there has been   work in cosmology that uses a notion of
semiclassical state \cite{luca}.

In this paper we  develop the semiclassical sector of quantum
cosmology. To do this we use a "polymer" quantization of
Friedmann-Robertson-Walker (FRW) cosmology in the
Arnowitt-Deser-Misner (ADM) canonical variables that was recently
presented in \cite{hw-cosm}. (For a related discussion using
connection-triad variables see Refs. \cite{mb1,mb2,abl}.)  We give a
construction of semiclassical states, and show that these states
have properties such as being peaked on a point in classical phase
space, and satisfying minimal uncertainty relations.

We then outline how these states can be used in applications.
First we discuss how to calculate quantum corrections to the
classical FRW dynamics by calculating  expectation values of
the quantum dynamical equations in those coherent states. There
are some options available for this, reflecting approaches one
can take to gravitational dynamics.

The second application concerns the question of the initial state of
the universe. The basic idea is to posit that the present state of
the universe is described by a semiclassical state (to be described
below), and then ask questions about the history of the universe by
evolving this state backward (or forward) in time. Or, put the other
way around, what quantum state when evolved for a sufficiently long
time leads to the state that we observe today, i.e. a semiclassical
state peaked on a flat  FRW cosmology with some matter content? This
requires a notion of time and its corresponding  true  Hamiltonian,
which we obtain  by fixing a time gauge.  This provides a
computational framework that allows one to compute the "initial
state" of the universe.

The paper is structured as follows: In section II we describe the
classical system. In section III we recall its quantization as
developed in \cite{hw-cosm}, with slight modifications to better
suit our goals here, and then introduce semiclassical  states. We
prove several properties which are physical requirements for a
semiclassical interpretation of these states. In the final section
we present an outline of two interesting applications of these
coherent states.

\section{Classical theory}

Our starting point is the ADM  Hamiltonian
action for general relativity minimally coupled to a massless
scalar field
\be S= \frac{1}{8\pi G}\int d^3x dt \left(\t{\pi}^{ab}\dot{q}_{ab}
+ p_\phi \dot{\phi} - N H - N^a C_a\right) \ee
where the Hamiltonian and diffeomorphism constraints are
\bea H &=& \frac{1}{\sqrt{q}}\left(\t{\pi}^{ab}\t{\pi}_{ab} -
\frac{1}{2}\t{\pi}^2 \right)+ \sqrt{q}(\Lambda - R(q))
+ 8\pi G(\frac{1}{2\sqrt{q}}p_\phi^2 + \frac{1}{2}\sqrt{q}q^{ab}\p_a\phi\p_b\phi) \\
C_a &=& D_c\t{\pi}^c_a + 8\pi G p_\phi\p_a \phi, \eea
where $\t{\pi}=\t{\pi}^{ab}q_{ab}$, $R(q)$ is the Ricci scalar of
the spatial metric $q_{ab}$, and $\Lambda$ is the cosmological
constant.

A reduction to the flat homogeneous isotropic case may be done by
writing  a parametrization for the canonical pair
$(q_{ab},\t{\pi}^{ab})$. For FRW cosmology a suitable choice is
\bea
q_{ab} &=& a^2(t) e_{ab} \\
\t{\pi}^{ab} &=& \frac{p_a(t)}{2a} e^{ab}
 \eea
where $e_{ab}$ is the flat Euclidean metric. Plugging this into
the ADM 3+1 action gives the reduced action
\be S^{\rm Red} = \int dt \frac{1}{8\pi G}\left(p_a \dot{a} +
p_\phi \dot{\phi} - N H_R \right) \ee
where the reduced Hamiltonian constraint, or equivalently the
Friedman equation in canonical coordinates, is
\be H_R = -\frac{3}{8} \frac{p_a^2}{|a|} + |a|^3\Lambda + 8\pi G
\frac{p_\phi^2}{2|a|^3} =0. \label{Hcons}
\ee
The fundamental Poisson bracket relations are
\be \{a,p_a\} = 8\pi G,\ \ \ \ \ \ \ \ \{\phi,p_\phi\} =1. \ee
The topology of the reduced phase space (for gravity and matter)
is $R^2\times R^2$. In $c=1$ units the gravitational phase space
variables each have dimension length.

The configuration and translation variables $a$ and
\be U_\lambda(p_a)={\rm exp}\left( i\lambda p_a/L\right) \ee
satisfy the algebra
\be \{a,U_\lambda\} = \frac{8\pi G}{L}\ i\lambda U_\lambda. \ee
As the classical limit for the coherent states constructed below
is obtained by sending a dimensionless parameter $t$ to zero (the
correct implementation of the textbook style "$\hbar \to 0$"
limit), it turns out to be useful to work with the dimensionless
variables $\t{a}=\frac{a}{L}$ and $\t{p}_a=\frac{p_a}{L}$, for
which the Poisson bracket becomes
\be \{\t{a},\t{U}_\lambda\}= \frac{8\pi G}{L^2}\ i\lambda
\t{U}_\lambda, \ee
where $\t{U}_\lambda = {\rm exp}(i\lambda \t{p}_a)$. From here
onwards we drop the tilde and use the dimensionless variables.
This is the basic bracket that will be realized as a commutator in
the quantum theory.

Another observable of interest is the inverse scale factor $1/a$,
which may be represented by classical identities of the type
\cite{thomas}
\be \frac{1}{|a|} = -\frac{4 L^4}{(8\pi G)^2\lambda^2}\
\left(U^*_\lambda \left\{U_\lambda,\sqrt{|a|} \right\}\right)^2.
\label{1/a} \ee
Such expressions are useful in that a representation of the
variables $a$ and $U_\lambda$ leads, via the right hand side, to a
realization of inverse scale factor and curvature operators that
are well-defined even on the state corresponding to the classical
singularity.

For a dynamical system with a
Hamiltonian constraint there are two ways to proceed to
quantization-- either via a reduced Hamiltonian obtained from a
time gauge-fixing, or by a Hamiltonian constraint operator. The
former has the advantage that it "solves" the problem of time at
the classical level by an explicit deparametrization, but it
leaves open the question of unitary equivalence of quantum
theories obtained from different time gauges. The latter is aimed
at obtaining fully gauge invariant states in which the problem of
time must still be addressed in some way before dynamical
processes can be described.

In vacuum gravity a gauge choice is a suitable function of the
canonical coordinates and momenta $t=f(q,\pi)$. The functions $f$
may be subdivided into three classes -- intrinsic, extrinsic or
mixed depending on whether $f$ depends on the 3-metric $q$, its
conjugate momentum $\pi$, or on both. If there is  matter
coupling, there is the additional possibility of choosing the time
(and other coordinate gauges) by choosing functions $f$ that
depend only on the matter variables. When a canonical gauge fixing
condition is used, the requirement that it be preserved in time
leads to equations that fix the lapse and shift functions.

For the covariant Einstein equations, gauge choices are made by
fixing directly the lapse and shift. For example the choice
$N=$constant is often made in cosmology. A canonical choice for
gravity coupled to a scalar field that has been much studied is
$t=\phi$ \cite{isham,gotay}, especially in relation to the problem
of curvature singularity avoidance. We exhibit here the canonical
gauge-fixing conditions corresponding to the  covariant choices
$N=1$ ("Hubble time") and $N=a(t)$ (conformal time), and derive
the corresponding reduced Hamiltonians. We do the same for the
gauge condition $\phi=t$.

Consider first the time gauge $t=ka^2/p_a$ where $k$ is a
constant. The requirement that this condition be preserved in time is
\be \dot{t} = 1= \left\{ \frac{ka^2}{p_a}, NH \right\},\ee
which gives (for $\Lambda =0$)
\be N= sgn(a) k^{-1} \left(-\frac{9}{8} - \frac{12 \pi G
p_\phi^2}{a^2p_a^2}\right)^{-1}.
\ee
Solving the Hamiltonian constraint strongly gives
\be \frac{8\pi G p_\phi^2}{a^2p_a^2} = \frac{3}{4}, \ee
so we get $N=-sgn(a) 4/9k$. Thus $k=-sgn(a)9/4$ gives $N=1$. The
main point here is the observation that Hubble time gauge corresponds
to a canonical time choice proportional to $a^2/p_a$.

The reduced Hamiltonian is proportional to the variable canonically conjugate
to the time choice, evaluated on the solution of the Hamiltonian
constraint. For the Hubble time gauge this is
\be h = \pm \frac{2}{3t} \sqrt{\frac{8\pi Gp_\phi^2}{3}}. \ee

Consider next the time gauge $t=ka/p_a$ where $k$ is again a
constant. This leads to the lapse function
\be N = sgn(a)\frac{a}{k}\ \left(-\frac{9}{8} - \frac{12 \pi G
p_\phi^2}{a^2p_a^2} \right)^{-1}, \ee
which gives $N=a$ for $k= -9/4$. This leads to the FRW metric
written in conformal time . The corresponding reduced Hamiltonian is
\be h=\pm \frac{1}{t} \sqrt{\frac{8\pi Gp_\phi^2}{3}}, \ee
which is proportional to the Hubble time Hamiltonian. This is an
accident of homogeneity and the fact that $ht \sim ap_a$ for both
of these time choices. Note also that the Hamiltonian gauge
conditions corresponding to these commonly used covariant gauge
choices are a mixture of the geometrodynamic coordinates and
momenta.

Finally for the $t=\phi$ gauge, the lapse function is
\be N =sgn(a) \frac{a^3}{8\pi Gp_\phi}, \ee
and the reduced Hamiltonian is
\be h=\pm \sqrt{\frac{3}{32\pi G}}\  ap_a, \ee
which is time independent.

\section{Quantization}

We summarize briefly the quantization  of this model presented in
\cite{hw-cosm}. The definition of basic variables used here is
slightly different in that  we use dimensionless phase space
variables for quantization, which leads to a dimensionless
parameter $t=(l_P/L)^2$ in the operator expressions ($l_P$ is the
Planck length and $L$ is an external scale). This parameter is
then utilized in the construction of semiclassical states.

The (kinematical) Hilbert space on which the basic variables are
realized has a basis given by the kets $|\mu\rangle \equiv
|exp(i\mu p_a)\rangle$, where the quantum numbers $\mu \in R$,
with the inner product
\be
\langle \mu|\nu\rangle = \delta_{\mu\nu}.
\ee
The basic variables are represented by
\bea
\hat{a}|\mu\rangle &=& 8\pi t \mu|\mu\rangle, \\
\hat{U}_\lambda |\mu\rangle &=& |\mu-\lambda\rangle, \label{ops}
\eea
which gives the commutator
\be
[\hat{a},\hat{U}_\lambda]=-8\pi t \lambda \hat{U}_\lambda.
\ee
The (kinematical) Hilbert space is not separable since, unlike the
Schr\"odinger representation, the inner product is such that
configuration variable eigenstates are normalizable. As a
consequence, the infinitesimal generators of translations, i.e.
operators corresponding to $p_a$, do not exist in this Hilbert
space. This is the essential difference from the Schrodinger
representation, and it leads to a fundamental inherent lattice
structure at the quantum level. The interested reader is referred to
\cite{hw-cosm} for more details, and to \cite{halvorson,polymer} for
other applications of this type of quantization.

With the representation (\ref{ops}) an inverse scale factor
operator is readily constructed using (\ref{1/a}). It is diagonal
in the basis with eigenvalue given by
\be \widehat{\frac{1}{|a|}}\ |\mu\rangle =\frac{1}{2\pi\lambda^2 t}\
\left( |\mu|^{1/2} - |\mu-\lambda|^{1/2} \right)^2 |\mu\rangle. \ee

Although the (kinematical) Hilbert space used in this quantization
is not separable, the dynamics selects a separable subspace, once an
initial state has been chosen. Thus all computations are naturally
restricted to separable subspaces, and this extends to the
semiclassical sector constructed below. An example is given by the
span of the vectors
\be |m\rangle \equiv |\mu_0 + m\mu \rangle, \label{subsp} \ee
where $m$ is an integer, and $\mu$ and $\mu_0$ are arbitrary real
numbers; $\mu_0$ may be viewed as the "origin" of  a lattice with
spacing given by $\mu$. In order to utilize this subspace, we must
work with operators that do not take us out of it. Since all
operators are constructed from $a$ and $U_\lambda$, this is
accomplished by working only with those $U_\lambda$'s adapted to the
subspace, ie. we must set $\lambda = \mu$. In the following we set
$\mu_0=0$ and work in the subspace $|m\lambda>$.

\section{Semiclassical states}
Coherent states for a particle in the Schr\"odinger representation
are of the form
\be
\psi \sim \frac{1}{\sqrt{t}}\ e^{-(x-x_0)^2/2t +ixp_0}.
\label{coherqm}
\ee
These are peaked at the classical phase space point $(x_0,p_0)$ --
in this sense they are semiclassical. The peaking properties
may be seen  by computing the expectation values of the $\hat{x}$
and $\hat{p}$ operators in this state. They also have the
additional property that they are eigenstates of the operator
$\hat{x}-i\hat{p}$.

We would like to construct semiclassical states for FRW cosmology
in the representation described above, motivated by the same
considerations. Such states have been discussed in loop quantum
gravity \cite{tw}, where the holonomy of a connection based on a
spatial loop is the analog of the translation variable
$U_\lambda$. In particular the case of the $U(1)$ gauge theory
coherent states discussed there is similar to what we require.

These considerations motivate the definition of states
\be |\alpha,\beta\rangle_{t,\lambda} =
\frac{1}{C}\sum_{m=-\infty}^\infty e^{-\frac{t}{2}(\lambda m)^2}
e^{m\lambda \alpha} e^{im\lambda \beta }|m\rangle. \label{coherb}
\ee
The normalization constant $C>0$ is given by the convergent sum
\be
C^2=\sum_{m=-\infty}^\infty e^{-t\lambda^2m^2} e^{2\alpha
\lambda m}. \label{norm}
\ee
 The real parameters
$\alpha$ and $\beta$ correspond to a classical configuration in
the same sense as the parameters $x_0,p_0$ in the state
(\ref{coherqm}), as we now show.

A first check is to verify that the states $(\ref{coherb})$ are
eigenstates of an operator analogous to the annihilation operator
$\hat{A}=\hat{x}+i\hat{p}$ for a Schrodinger particle. However,
since the momentum operator is not directly represented in this
quantization, the closest we have is the exponential $e^A\equiv
e^{x+ip}$, which is represented by the operator
\be
e^{\gamma \hat{a}}\ \hat{U}_\lambda(\hat{p}_a),
\ee
where the parameter $\gamma$ is determined by the condition that
the state (\ref{coherb}) is an eigenstate of it. It is
straightforward to verify that
\be
e^{(\lambda/8\pi)\hat{a}}\ \hat{U}_\lambda
|\alpha,\beta\rangle_{t,\lambda} = e^{t\lambda^2/2}
e^{\lambda(\alpha+i\beta)} |\alpha,\beta\rangle_{t,\lambda}.
\ee
This result suggests that the expectation values of operators
$\hat{O}(\hat{a},\hat{U}_\gamma)$ are peaked at the corresponding
classical phase space functions $O(a,p_a)$ in the limit
$t\rightarrow 0$. That this is in fact the case is established by
direct calculation of expectation values. The limit $t\rightarrow
0$ requires use of the Poisson re-summation formula
\be \sum_{m=-\infty}^\infty f(s m) =
\frac{2\pi}{s}\sum_{m=-\infty}^\infty\bar{f}\left(\frac{2\pi
m}{s}\right), \label{pres} \ee
where $s$ is a real parameter, $f$ is function on the real
line, and $\bar{f}$ is its Fourier transform
\be \bar{f}(k)= \int_{-\infty}^\infty dx\   f(x)e^{-ikx}. \ee

As illustrative examples, let us compute in the state
(\ref{coherb}) the expectation value of $\hat{a}$,
$\hat{U}_\lambda$, and of an expression for the momentum operator.
The latter in this quantization is represented by
\be
\hat{p}^\lambda_a \equiv \frac{1}{2i\lambda}\
\left(\hat{U}_\lambda - \hat{U}^\dagger_\lambda\right).
\label{mom}
\ee
Let us first note that the normalization constant for the
semiclassical states (\ref{norm}) may be rewritten using
(\ref{pres}) as
\be
C^2= \sqrt{\frac{\pi}{\lambda^2 t}}\ e^{\alpha^2/t}\left( 1+
2\sum_{m\ne 0}\cos\left( \frac{2\pi m\alpha}{\lambda t} \right)
e^{-\pi^2m^2/t\lambda^2} \right).
\ee
This form of the result facilitates taking the $t\rightarrow 0$
limit since the terms in the sum gets damped to zero. Using this
we obtain
\bea \langle\hat{a}\rangle &=&  8\pi\alpha\ \left(\frac{1}{1+
2\sum_{m\ne 0}\cos\left[ \frac{2\pi m\alpha}{\lambda t} \right]
e^{-\pi^2m^2/t\lambda^2}}\right),\label{expeca}\\
\langle \hat{U}_\lambda \rangle &=&  e^{i\lambda \beta}\
e^{-t\lambda^2/4}\ \left(\frac{  1+ 2\sum_{m\ne 0}\cos\left[
\frac{2\pi m\alpha}{\lambda t}\left(1+
\frac{t\lambda}{2\alpha}\right) \right] e^{-\pi^2m^2/t\lambda^2}
}{  1+ 2\sum_{m\ne 0}\cos\left[ \frac{2\pi m\alpha}{\lambda t}
\right] e^{-\pi^2m^2/t\lambda^2} }\right). \label{expecU}
\eea
Note that these expressions have the limits expected of
semiclassical states:
\bea
\lim_{t\rightarrow 0}\ \langle \hat{a}\rangle &=& 8\pi\alpha, \label{expa}\\
\lim_{t\rightarrow 0}\ \langle \hat{U}_\lambda \rangle &=&
e^{i\lambda\beta}. \label{expU}
\eea

Equation (\ref{expecU}) together with the definition (\ref{mom})
gives the expectation value
\be
\langle \hat{p}_a^\lambda\rangle =
\frac{\sin(\beta\lambda)}{\lambda}\ e^{-t\lambda^2/4}\
\left(\frac{ 1+ 2\sum_{m\ne 0}\cos\left[ \frac{2\pi
m\alpha}{\lambda t}\left(1+ \frac{t\lambda}{2\alpha}\right)
\right] e^{-\pi^2m^2/t\lambda^2} }{  1+ 2\sum_{m\ne 0}\cos\left[
\frac{2\pi m\alpha}{\lambda t} \right]
e^{-\pi^2m^2/t\lambda^2}}\right),
\ee
where the Poisson re-summation formula has been used in the last
step. This formula has the limits
\bea \lim_{t\rightarrow 0}\ \langle \hat{p}_a^\lambda\rangle
&=& \sin(\beta\lambda)/\lambda, \\
 \lim_{\lambda\rightarrow 0}\ \langle \hat{p}_a^\lambda\rangle
&=& \beta.
 \eea
The first shows that the semiclassical state on the lattice is
peaked at the corresponding phase space value. The second shows that
the continuum limit of the momentum expectation value has the
appropriate peaked value in this state, even though only the
translation operators exist in the representation we are using for
the quantum theory \footnote{After this work was completed, we
became aware of ref. \cite{fr} where there is a related discussion
of an approximate momentum operator in this representation.}.

It is also possible to define $\lambda$ dependent creation and annihilation operators
in this quantization via
\be \hat{A}^\lambda \equiv \hat{a} - i \hat{p}_a^\lambda \ee and
its adjoint.
From the above result for the expectation value of
$\hat{p}_a^\lambda$ it follows that
\be \lim_{t\rightarrow 0}\ \langle  \hat{A}^\lambda \rangle =
\alpha +i\beta,
\ee
and hence that the semiclassical state with $\alpha = \beta =0$
may be compared to the usual oscillator vacuum in this alternative
quantization. This "vacuum" $|\alpha =0,\beta =0\rangle $ may be
viewed as a "collective" state in the sense that it is an infinite
linear combination of suitably weighed elementary states
$|m\rangle$. The states resulting from repeated action of
$A^{\lambda\dagger}$ on this vacuum similarly provide a
correspondence with the excited oscillator states. This idea may
also be applied to a related quantization of the scalar field
\cite{als}, and the gravity scalar field model in spherical
symmetry \cite{hwbh} to gain more insight into how the usual
background dependent Fock quantization of the scalar field is
related to the present one.

It is also possible to see that the wave function corresponding
to the state (\ref{coherb}) is peaked in the same way as the one
for the oscillator. Recall that the basis elements in which we
write the semiclassical state  are configuration eigenstates with
wave function $e^{-i\lambda m p}$. Therefore the momentum space
wave function corresponding to the state (\ref{coherb}) is
\be
\psi_{(\alpha,\beta)}(p) = \frac{1}{C}\
\sum_{m=-\infty}^\infty e^{-\frac{t}{2}(\lambda m)^2} e^{m\lambda
\alpha} e^{im\lambda (\beta-p)}.
\ee
This gives the momentum space probability distribution
\be \frac{1}{C^2}\ |\psi_{(\alpha,\beta)}(p)|^2  =
\sqrt{\frac{4\pi}{\lambda^2t}}\ e^{-(p-\beta)^2/2t}\
\frac{\left|1+ 2\sum_{m\ne 0}\cos\left( \frac{2\pi
m\alpha}{\lambda t} \right) e^{-\pi^2m^2/t\lambda^2-2\pi
m(p-\beta)/\lambda t}\right|^2} {1+2\sum_{m\ne 0}\cos\left(
\frac{2\pi m\alpha}{\lambda t} \right) e^{-\pi^2m^2/t\lambda^2}}.
\label{prob-p}
\ee
It is evident from this expression that in the limit $t\rightarrow
0$ the distribution is peaked at the momentum value $\beta$, just
as the position space oscillator wave function (\ref{coherqm}) is
peaked at $x_0$. This is because the sums in the numerator and
denominator are damped to zero in this limit.

So far we have verified that the peaking property of the
expectation value holds for the basic phase space variables $a$
and  $U_\lambda(p_a)$.
This is expected to be true also for any function of the basic
variables. For example, the explicit calculation for the
expectation value of the inverse scale factor (\ref{1/a}) gives
\be \lim_{t\rightarrow 0}\left\langle\widehat{\frac{1}{a}}
\right\rangle = \frac{1}{8\pi \alpha},
\ee
which is the inverse of $\langle \hat{a} \rangle$ (\ref{expa}) in
this limit.

All the above properties establish that the states defined in eqn.
(\ref{coherb}) have the required semiclassical properties.

\section{Applications and Discussion}
In this section we propose two  applications of
semiclassical states to cosmology. The first concerns computing
quantum corrections to classical dynamics and makes use of the Heisenberg
interpretation. The second concerns implementing an idea to obtain the wave
function at early times by evolving a semiclassical state backward in time,
and uses the Schrodinger representation. Their implemention will appear
elsewhere.

\subsection{Quantum corrections to classical dynamics}

One of the expectations from a quantum theory of gravity is that
it provide a mechanism for the emergence of a classical spacetime
in an appropriate limit, and a procedure for computing
quantum corrections to classical equations. An immediate application
would be to cosmology, which is perhaps the only arena where a quantum
gravity theory may be testable.

A possible approach for computing quantum corrections to classical
cosmological equations is suggested by the peaking results for semiclassical
states proven above. The basic idea is to obtain the Heisenberg equations
of motion for the relevant observables, and compute the expectation values of
the commutator terms in the semiclassical states. In the $t\rightarrow 0$
limit the resulting equation would give the classical equations because to the results
of the last section.

There are two ways that this idea can be implemented -- with or without
a time gauge fixing. In the former case the Hamiltonian $h$ corresponding to
the time gauge fixing is derived at the classical level and
converted to an operator. The quantum corrected equations are then postulated
to be
\be \dot{a} = \langle\alpha,\beta |\left[\hat{a},\hat{h}\right]
|\alpha,\beta\rangle \ee
for the scale factor, with similar equations for the scalar field and the
conjugate momenta. The right hand side may be expanded in powers of the parameter
$t$ to give the classical term and its corrections order by order.

If working without a time gauge fixing, the Hamiltonian constraint
operator would be used with an arbitrary lapse function to obtain
the evolution equation. This requires  a definition  of the
Hamiltonian constraint operator $\hat{H}_R$ coming from the
classical expression (\ref{Hcons}). The square of the momentum in
this constraint may be  realized by the operator
\be
\left(\hat{p}^\lambda_a\right)^2 = \frac{1}{\lambda^2}\
\left(2-\hat{U}_\lambda-\hat{U}_\lambda^\dagger. \right)
\ee
This, together with the operator corresponding to  $1/|a|$
given in Eqn. (\ref{1/a}) gives an expression for the first term in the
Hamiltonian constraint operator (with a choice of operator
ordering).

The semiclassical state peaked on a classical solution of the
constraint satisfies
\be \lim_{t\rightarrow 0} \langle \hat{H}_R \rangle =0 + {\cal
O}(t^\sigma), \ee
where the power $\sigma >0$ of the first quantum correction to
expectation value  may depend on the choice of operator ordering
in the $p_a^2/a$ factor in the constraint (\ref{Hcons}). Without
the limit, this equation gives quantum corrections to the Friedman
equation. Similarly, corrections to the Hamiltonian evolution are obtained by
computing the right hand sides of lapse dependent equations such as
\be \dot{a} = N \langle\alpha,\beta
|\left[\hat{a},\hat{H}_R\right] |\alpha,\beta\rangle.
\ee

We emphasize that this procedure is quite different from what is usually called "the
semiclassical approximation" in quantum gravity, which treats gravity classically and matter quantum
mechanically. The central difference is that here the matter and gravity variables are
treated at the same level -- the full state used to compute quantum corrections is the
tensor product of the matter and gravity semiclassical states. It is however possible,
and quite straightforward to obtain this usual and more limited approximation from our
more general procedure by taking the $t\rightarrow 0$ limit in the expectation values
for only those term that contain the gravitational variables. This effectively makes
gravity variables classical, with the matter and interaction parts receiving the $t$
dependent quantum corrections, with an implicit choice of "vacuum" defined by the
matter semiclassical state.

\subsection{Initial state of the Universe}

Semiclassical states may be used as a "present time condition" for
the cosmological state of the Universe. This is a reasonable
assumption because observations suggest that an FRW model provides
a good large scale description. This state may be evolved into the
past or the future using the Hamiltonian operator obtained by a
time gauge fixing. It is apparent from the form of the Hamiltonian
that such evolution leads, after some time steps, to a new state that
is not of the form (\ref{coherb}). An initial state can be tracked to
early times by following the evolution of the probability density
(\ref{prob-p}).

It is perhaps easiest to implement this procedure using the time
independent Hamiltonian obtained from the $\phi=t$ gauge. After
this gauge fixing the canonical variables are the pair $(a,p_a)$
with Hamiltonian $h\sim ap_a$. A time step evolution of an initial
state using a simple scheme such as
\be \psi (t+\Delta t)= \left(I + i\hbar \Delta t\
\hat{h}\right)\psi(t) \ee
may be implemented numerically to see how the state evolves to the
past and future, and also to obtain an idea of the degree of coherence
that is retained by evolution.

Although such evolution is unitary by construction, numerical implementation
restores it only up to some order in the time step $\Delta t$. For example for
the simple explicit scheme given above, unitarity is not exact with violations of
order $\Delta t^2$. There are known implicit schemes whose unitary behaviour is much
better. An example is provided by a modified Crank-Nicholson method where the
Schrodinger equation is discretised as
\be \frac{i}{\Delta t}\  \left[\psi(t+\Delta t) - \psi(t)\right] =
\frac{\hat{h}}{2}\left[\psi(t+\Delta t)+\psi(t)\right]. \ee
This time stepping scheme remains useful if the gauge fixing is such
that the Hamiltonian has explicit time dependence.

There are other physical situations where  these semiclassical
states may be used in cosmology. One of these is the question of
quantum gravity corrections to the spectrum of density
perturbations. There has been an initial exploration of this
question without coherent states \cite{os}, where a quantum gravity
corrected  FRW scalar wave equation  is obtained by replacing
inverse scale factor terms by the eigenvalue of the corresponding
operator (\ref{1/a}) in a basis state. In the energy regime where
this calculation is normally done, spacetime is approximately
classical. Therefore it would be interesting to do such a
calculation with the expectation value taken in the appropriate
semiclassical state, and expanded to the desired order in the Planck
length. This would give controlled corrections to the usual
quantum-fields-on-a-classical-background semiclassical
approximation. Work on developing these applications is in progress.

\bigskip
\noindent{\bf Acknowledgements}: This work was supported in part
by the Natural Science and Engineering Research Council of Canada.

%\end{multicols}

\end{document}